# Aerial Platform Design Options for a Life-Finding Mission at Venus

Weston P. Buchanan [1,*], Maxim de Jong [2], Rachana Agrawal [1], Janusz J. Petkowski [3], Archit Arora [1], Sarag J. Saikia [4], Sara Seager [3], James Longuski [1] and the Venus Life Finder Mission Team

[1] School of Aeronautics and Astronautics, Purdue University, West Lafayette, IN 47907, USA; racha.agrawal.04@gmail.com (R.A.); arora31@purdue.edu (A.A.); longuski@purdue.edu (J.L.)
[2] Thin Red Line Aerospace Ltd., Chilliwack, BC V2R 5M3, Canada; maxim@thin-red-line.com
[3] Department of Earth, Atmospheric, and Planetary Sciences, Massachusetts Institute of Technology, Cambridge, MA 02139, USA; jp2ch@virginia.edu (J.J.P); seager@mit.edu (S.S.)
[4] Spacefaring Technologies Pvt. Ltd., Bengaluru, 560049 India; saragjs@gmail.com
* Correspondence: buchanaw@purdue.edu

**Abstract:** Mounting evidence of chemical disequilibria in the Venusian atmosphere has heightened interest in the search for life within the planet's cloud decks. Balloon systems are currently considered to be the superior class of aerial platform for extended atmospheric sampling within the clouds, providing the highest ratio of science return to risk. Balloon-based aerial platform designs depend heavily on payload mass and target altitudes. We present options for constant- and variable-altitude balloon systems designed to carry out science operations inside the Venusian cloud decks. The Venus Life Finder (VLF) mission study proposes a series of missions that require extended in situ analysis of Venus cloud material. We provide an overview of a representative mission architecture, as well as gondola designs to accommodate a VLF instrument suite. Current architecture asserts a launch date of 30 July 2026, which would place an orbiter and entry vehicle at Venus as early as November 29 of that same year.

## 1. A Brief History of Atmospheric Probes at Venus

For decades, scientists have considered the possibility that life exists in the clouds of Venus [1–4]. Since Mariner 2 flew by the planet in 1962 [5], more than a dozen probes have plunged into the planet's cloud decks [6]. While scientists long wondered whether a habitable surface was shrouded beneath the thick Venusian atmosphere, closer scientific inspection quickly revealed quite the opposite: a dense, $CO_2$ lower atmosphere underlain by the hottest surface of any planetary body in the solar system [7–9].

Nevertheless, scientific interest in the inhospitable world persisted. It was suspected that Venus and Earth shared similar early histories [10], so exactly why Venus followed such a different developmental path became a topic of investigation. The Soviet Union established the Venera program, launching more than a dozen probes into the Venusian clouds to collect data. Notable achievements by the Soviet Union include the insertion of the first atmospheric probe into the Venusian atmosphere [6]; the first successful landing on the surface of Venus; the first surface images returned from Venus; the first in situ analysis of Venus surface material; and the insertion of the first buoyant platform into the Venusian (or any extraterrestrial) atmosphere.

Meanwhile, the United States focused its efforts on a single mission, Pioneer Venus. The mission consisted of an orbiter and a multiprobe bus carrying four descent probes (with the bus itself acting as a fifth probe that would burn up high in the atmosphere). All four descent probes relayed atmospheric data directly to Earth before impacting the surface. While none of the probes were designed to survive impact, one of them did, and even transmitted data for 67 min after striking the surface. The orbiter also completed its mission of gathering atmospheric, image, and other physical data—and even survived for well over a year beyond the end of the nominal mission [5].

In preparation for the Pioneer Venus mission, NASA considered a large buoyant platform but settled on descent probes due to their reduced complexity and cost [5]. Despite the many probes that have showered the planet over the decades, there has been only one balloon mission to Venus. In 1985, the Soviet Union successfully placed two balloons into the cloud decks. The VeGa balloons entered the atmosphere within a few degrees of the equator and drifted with the zonal winds for approximately two days as they were monitored by 20 radio observatories all over Earth [11].

At an altitude of 53 km, the balloons traveled at average speeds of nearly 70 m/s, covering distances exceeding 11,000 km before losing contact [12]. The identical balloon systems were minimal in design. The combined envelope, helium, and tether weighed 14.6 kg, and the gondola added just 6.9 kg [13].

In spite of its conservative allocation, the instrument suite within the gondola measured a variety of information. Among the instruments were temperature and pressure sensors, as well as an anemometer for measuring vertical wind speed. A telemetry package measured the balloon's trajectory and the local wind speed. A nephelometer measured cloud particle density and size, while a light sensor measured ambient illumination [13,14].

Despite VeGa's status as the only balloon mission ever flown within the Venusian atmosphere, several such concepts have been designed over the decades, including the Venus Climate Mission (VCM) [15], European Venus Explorer (EVE) [16], Balloon Experiment at Venus (BEV) [17], Buoyant Venus Station (BVS) [18], and the 2020 Venus Flagship Mission (VFM) [19]. We summarize them briefly for context as follows.

The VCM, a planetary decadal study by NASA, deploys a helium superpressure balloon with a float altitude of 55.5 km. The VCM balloon carries a mini probe and two dropsondes that are deployed at different times over a nominal operational time of 21 days [15]. The EVE, a mission proposal to the European Space Agency, consists of a helium superpressure balloon that supports a gondola housing an array of instruments. At an altitude of 55 km, the EVE science payload performs atmospheric measurements over a period of 10 days [16]. NASA JPL's BEV concept consists of a reversible fluid altitude control balloon supporting a gondola and instruments. In addition to measuring temperature, pressure, and wind direction, the BEV instrument suite gathers optical navigation data of the Venusian surface from below the clouds over a matter of days [17]. The BVS mission features a hydrogen superpressure balloon supporting a gondola, as well as two dropsondes. Over a one-week period, the BVS instruments return atmospheric data including altitude profiles [18]. The VFM proposes a pumped-helium altitude control balloon, which cycles between 52–62 km over a period of 60 days. The science instruments housed by the VFM gondola return a variety of atmospheric, image, and magnetic field data [19].

More recently, NASA selected DAVINCI, an atmospheric probe mission that will investigate various properties of the Venusian atmosphere. The instrument suite will be accommodated by a descent probe fitted with a parachute. Atmospheric inlets protruding from the descent probe will be heated in order to vaporize ingested cloud material—a significant difference from the VLF mission, which seeks to analyze intact cloud particles for astrobiological purposes [20].

## 2. Aerial Vehicle Selection

While the VeGa balloons returned valuable meteorological data, the VLF mission seeks to measure characteristics of cloud material that appear in concentrations much lower than could be detected by previous instrumentation—for example, measuring trace atmospheric gases to ppb-level abundance. Therefore, the capability of extended in situ analysis of multiple samples of cloud material becomes far more valuable. The challenge of choosing an aerial vehicle for atmospheric sampling at Venus is to balance the science return the vehicle can deliver with the risk incurred as a consequence of choosing that vehicle. For a given instrument suite, science return increases with the amount of time afforded within the cloud decks. Holding this time constant, science return tends to vary in proportion to mass, power, and volume allowances (with generous allowances permitting faster data rates, more populous instrument suites, and greater redundancy). Risk increases with the complexity of the vehicle.

Descent probes spend very little time in the cloud decks. For example, after entering the clouds, it was only about 13 min before the Pioneer Venus Large Probe—transmitting at 128 bits/sec [21] and fitted with a parachute that opened at around 67 km—plunged through the bottom of the clouds (~48 km) and into the lower atmosphere. The three smaller probes of the same mission, which only transmitted at 64 bits/sec during the normal descent phase [21], did not have parachutes and traversed the same distance in less than 5 min [5].

Moreover, the latitudinal and longitudinal variation of a descent probe's trajectory is limited to the extent by which the probe drifts under the influence of zonal winds throughout its descent.

While Pioneer Venus offered a partial solution to this problem by inserting multiple probes into the Venusian atmosphere at different locations, continuous latitude-longitude profiles of atmospheric measurements require a vehicle capable of circumnavigating the planet.

Cutts et al. state that a hybrid airship yields high science return, but also dramatically increases size and complexity [22]. An enhanced-range (20 km) variable-altitude balloon (VAB) offers slightly less science return but imposes less risk. However, the development of such a balloon poses distinct challenges. The VAB proposed in NASA's 2020 Venus Flagship Mission Study is designed for no more than 14 km of altitude variation (50–64 km) [19]. A constant altitude balloon (CAB) lowers risk further, but at the additional expense of science return. A solar plane offers science return comparable to that of a CAB but imposes greater risk due to its complexity and technological maturity [22].

Because of the balance offered between mission risk and science return, a balloon is widely considered to be the most desirable vehicle for a prolonged in situ investigation of the Venusian atmosphere given current technological constraints. We consider designs for both constant- and variable-altitude balloons. While a VAB is more valuable than a CAB in terms of science return, the latter imposes less risk, is less costly, and may be required as a precursor for a VAB mission. Moreover, should the altitude band of interest be constrained to within a few km, the science return of the VAB and CAB are virtually equivalent. This is because the CAB's altitude varies passively due to updrafts and downdrafts over the duration of the mission. It is only when the range of targeted altitudes extends beyond a narrow band that the VAB convincingly surpasses the CAB.

**3. Altitudes of Interest**

The altitudes of astrobiological interest to the VLF team currently span the entirety of the cloud decks (~47.5 km to ~70 km above the surface). Many intriguing in situ observations of Venus' clouds have never been fully explained (see Figure 1). Such anomalous observations include: the unknown UV absorber in the upper cloud layers (57–70 km) [4]; the presence of non-volatile elements, including phosphorus and iron (in the middle and lower clouds, 47.5–56 km) [23,24]; the anomalous non-spherical Mode 3 particles with unknown composition (predominantly in the lower clouds 47.5–50.5 km) [25]; the presence of several gases with abundances out of thermodynamic equilibrium, including $SO_2$, $H_2O$, $O_2$, $H_2S$, and tentatively, $PH_3$ and $NH_3$ (detected predominantly in the mid-low cloud decks at 47.5–56 km, and in the haze layer below the clouds at 31–47.5 km) [26–30].

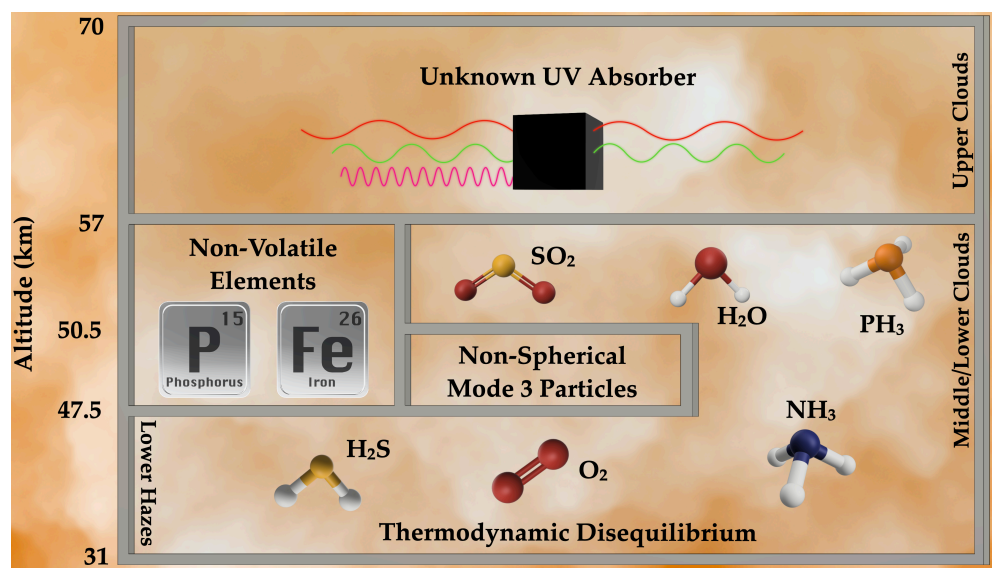

**Figure 1.** Unexplained Venus atmospheric anomalies in the clouds that could be targets of a dedicated prolonged scientific investigation by an aerial platform.

For a more in-depth overview of the Venusian cloud anomalies that motivate the VLF missions, please see [31,32]. For a discussion on the current status of the Venusian $PH_3$ discovery, please see [33]. Nearly all of these observations might be the result of life's activities, though, alone, life may not be required to explain any of them. These anomalies, both individual and taken together, are significant scientific motivators to explore the cloud deck altitudes and return to Venus with in situ observations.

### 4. Mission Overview

The baseline mission consists of an entry probe and an orbiter launched in a stacked configuration on 30 July 2026. The orbiter acts as a carrier spacecraft for the entry probe during the 122-day interplanetary cruise phase. The aerial platform is stowed inside the entry probe. As the spacecraft approaches Venus, the orbiter releases the entry probe and performs a deflecting maneuver to intercept the target B-plane at a periapsis altitude of 1000 km and insert into a 6-h orbit. The orbiter functions as a communication relay between the balloon and the Earth station. The entry probe deploys the balloon and inflation system in the atmosphere of Venus. Primary mission events are depicted in Figure 2. We consider three variations of this mission architecture.

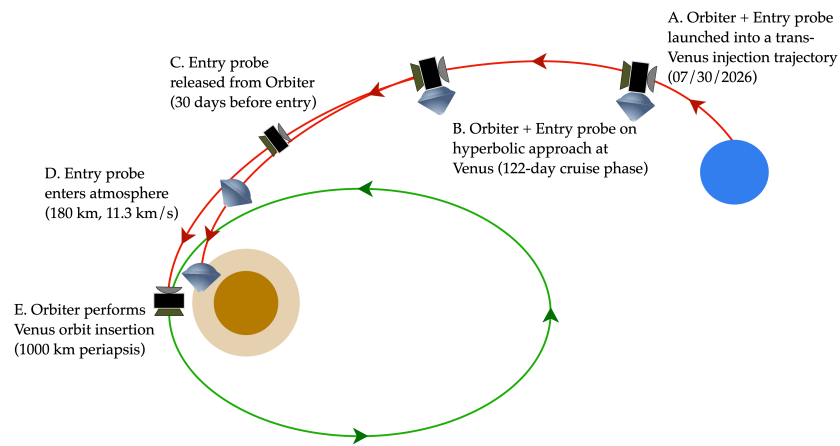

**Figure 2.** Baseline mission architecture for a 2026 launch.

The first of the three mission variations considered—the Habitability Mission—is designed to search for evidence of life in the Venusian clouds, measure habitability indicators, and characterize aspects of Venusian cloud droplets and aerosols that might be associated with life. The mission concept consists of an aerial platform that floats at an altitude of 52 km to perform science operations. The nominal duration of the mission is one week. The most conservative of the concepts considered, the Habitability Mission features roughly half as many instruments as its alternatives [34].

The most ambitious of the concepts herein considered, the Venus Airborne Investigation of Habitability and Life (VAIHL) Mission accomplishes all of the goals of the Habitability Mission and has the additional capability of directly searching for morphological indicators of life. The mission consists of a VAB that floats between the altitudes of 45 km and 52 km and is able to pause at altitudes of interest and perform science operations. The nominal duration of the mission is 30 days [34].

The VAIHL Lite Mission is similar to the VAIHL Mission but with less aerial platform mobility and a smaller, less diverse instrument suite. The float altitude of the VAIHL Lite CAB varies only passively over the course of the mission. Without a mass spectrometer or microscope, the VAIHL Lite instruments have a reduced capacity to identify evidence of life and characterize aspects of droplets and aerosols that might be associated with life (e.g., the VAIHL Lite suite cannot detect and characterize morphological indicators of life or determine the amount of water vapor in the clouds) [34].

### 5. Balloon System Options

The VAB and CAB developed by Thin Red Line Aerospace [35] share several design characteristics. Both are fitted with pressure restraint tendons made with Zylon © PBO fiber. To protect against the acidic environment within the cloud deck, the envelopes have a fluoropolymer membrane as an acid-resistant exterior cloak. The envelopes also incorporate an innovative five-layer metallized gas barrier laminate for long-term retention of the balloon's lifting gas.

Planar fabrication and packaging reduce precious stowage volume requirements, as well as the risk of trauma and leaks during transit and deployment (see Figure 3). Though the size of the inflation system changes between the VAB and CAB, both are composed of several cylindrical Luxfer T41A helium tanks arranged about the equator of the gondola.

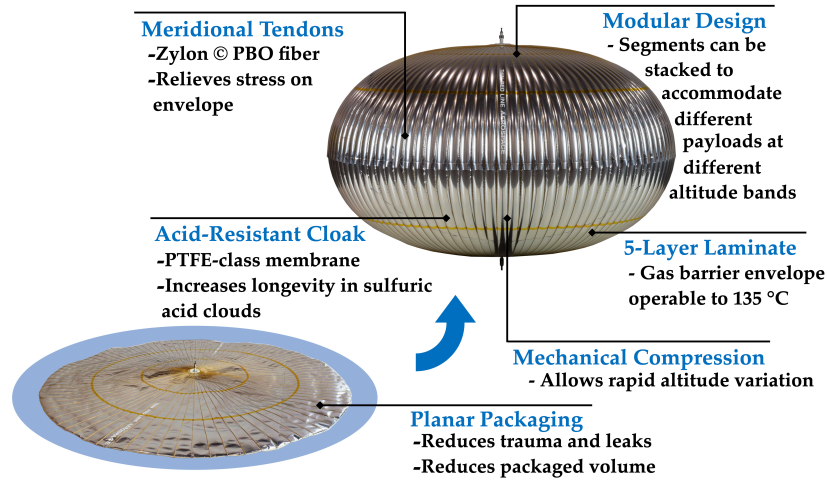

**Figure 3.** A Thin Red Line balloon segment.

Several mechanisms of altitude control exist. Pumped-helium balloons—widely favored due to their relatively low mass and power requirements—transfer helium between an inner chamber and outer envelope. The inner chamber is of constant volume while the volume of the outer envelope varies to maintain equilibrium with the ambient atmosphere. One such balloon was proposed in [19]. Reversible fluid altitude control balloons rely on the vertically striated temperatures on Venus to vaporize buoyancy fluids at low altitudes and subsequently condense them at high altitudes, producing positive and negative buoyancy, respectively [17]. Pumping atmosphere into and out of the envelope as ballast is not favored in the context of a Venus mission concept as the harmful effects of ambient sulfuric acid would necessitate the inclusion of acid-resistant materials on both the exterior and interior of the balloon [19].

The variable-float-altitude (VFA) balloon circumvents the problem of acidic ballast through its use of reserve gas carried on the gondola. To change altitude, the balloon uses an innovative combination of (a) conventional ballast drop and controlled gas venting, and (b) strategic use of heating and cooling associated with the Venus diurnal cycles. The VFA concept all but eliminates the complexity of a variable-altitude balloon (VAB) that incorporates active mechanical or pump-based density control systems, while facilitating a limited yet still reasonable number of altitude cycles, depending on the scope of intended science objectives [36].

The VAB varies its altitude using mechanical compression. A cable is passed axially through a stack of 6–12 balloon segments and fixed to a motor at the base of the column. Ascent and descent of the VAB is initiated and maintained by means of lifting gas density modulation through motorized mechanical compression, while the balloon's accordion-like envelope simultaneously allows the lifting gas volume to adapt to atmospheric density through a significant range of trajectory altitudes (see Figure 4). This is in contrast to the CAB, which maintains a constant volume for the duration of the mission.

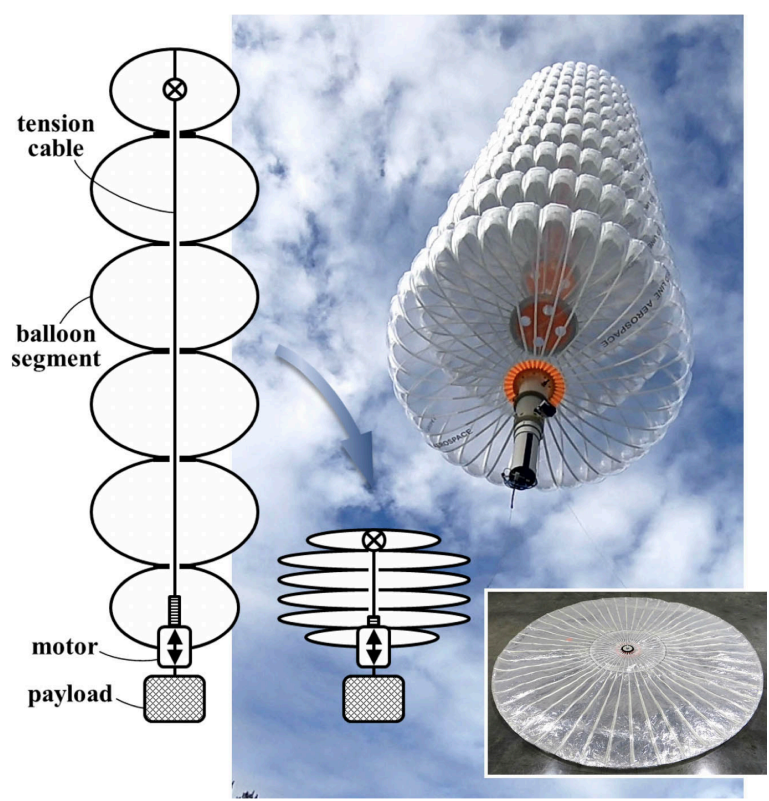

**Figure 4.** An illustration of mechanical compression of the VAB envelope. Pressure restraint tendons are visible in the top-right. Planar packaging configuration is visible in the bottom-right.

Mechanical compression maintains superpressure for the duration of the mission. This is beneficial as it provides instantaneous altitude control while increasing the robustness of the balloon. Moreover, a turgid envelope yields a more aerodynamic drag profile. Specific to the mechanical compression balloon, a continuously taut tension cable reduces the risk of "line dig", in which a cable is loosely wound on a winch drum and subsequently highly loaded.

The mechanical compression system delivers maximum ascent and descent rates as high as 8 m/s—the latter measuring an order of magnitude higher than that of a pumped-helium balloon (e.g., ~0.14 m/s is proposed as the maximum descent rate in [19]). Because of the superrotating atmosphere of Venus, aerial platforms drift considerably during ascent and descent. Greater ascent and descent rates reduce the amount of horizontal drift undergone during altitude excursions and increase the maximum number of excursions that can be performed during a given mission. This better enables the aerial vehicle to accommodate science operations related to vertical atmospheric profiling.

Figures 5 and 6 show the total in-flight mass and envelope volume as functions of payload mass for four balloon system designs, each corresponding to a different altitude (or altitude band).

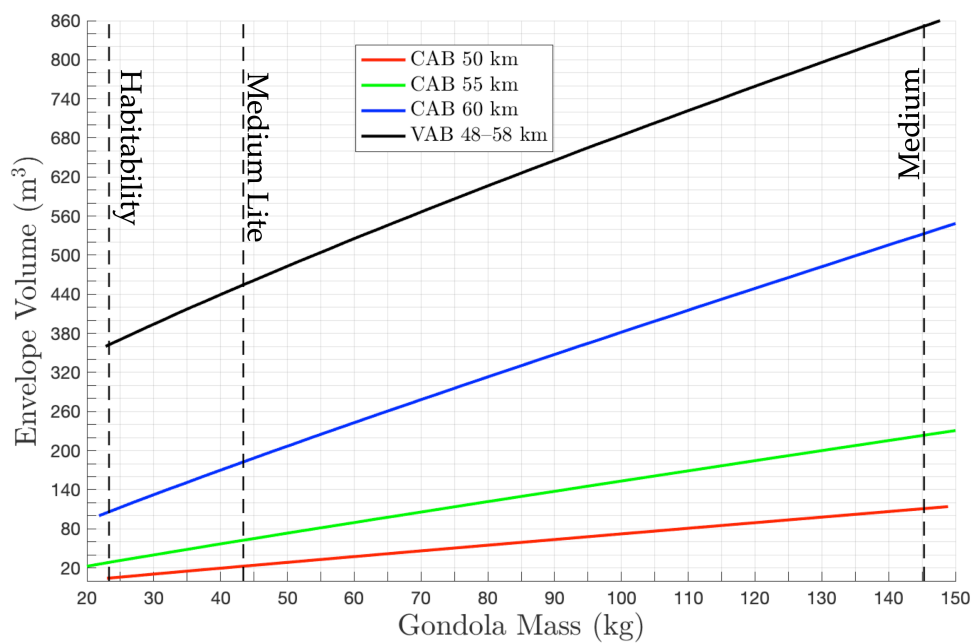

**Figure 5.** In-flight mass as a function of gondola mass for various Thin Red Line balloons based on calculations provided by Thin Red Line Aerospace Ltd., Chilliwack, BC, Canada. The x-axis represents gondola mass in kg. The y-axis represents total in-flight mass in kg. For CABs between 50 km and 60 km, the mass penalty for increasing float altitude is on the order of 1 kg/km. For every kg of mass added to the gondola, we see approximately 1.5 kg added to the in-flight mass. A 48–58 km VAB has an in-flight mass at least 160 kg greater than the heaviest CAB accommodating the same payload.

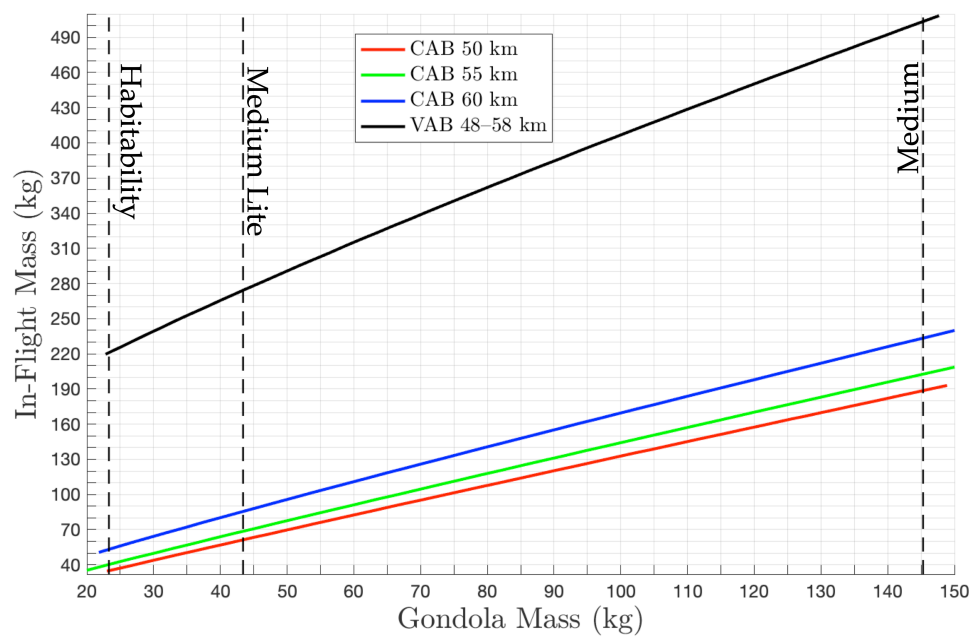

**Figure 6.** Envelope volume as a function of gondola mass for various Thin Red Line balloons based on calculations provided by Thin Red Line Aerospace Ltd. The x-axis represents gondola mass in kg. The y-axis represents envelope volume in m$^3$. On average, for CABs between 50 km and 60 km, the envelope volume more than doubles for every 5 km increase in float altitude. For every kg of mass added to the gondola, the increase in envelope volume is on the order of 1 m$^3$. A 48–58 km VAB has an envelope volume at least 240 m$^3$ greater than the largest CAB accommodating the same payload.

The CAB design results in a much lighter system than that of the VAB for a given float altitude. Again, we see that mass and volume requirements for the CAB increase with float altitude, but with the requirements themselves being several times less than those of the VAB. This difference is partly because the envelope of the CAB is much smaller than that of the VAB. The other significant difference is that the CAB does not include a flight control system.

In our analysis, we considered multiple scaled versions of the design for the VAB, each corresponding to a unique altitude band. In general, for an altitude band of a given width, mass and volume increase significantly as the upper bound of the altitude band increases. Mass and volume requirements decrease significantly with the width of the band.

## 6. Gondola Design

The gondolas for the VAIHL, VAIHL Lite, and Habitability Mission concepts are spherical pressure vessels machined from two titanium hemispheres of 2–6 mm thickness. They each contain a single beryllium shelf, which supports a science payload, batteries, a power distribution system, a command and data handling subsystem, and a communications subsystem. The shelf, which also acts as a heat sink, is coated on both sides by a 6 mm layer of sodium silicate. The thickness of the shelf is the same as that of the hull and varies depending on the mission concept. The inner walls of the pressure vessel are lined by a 1 mm thick Kapton blanket. A crossed dipole antenna protrudes from the aft hemisphere.

While the instrument suite of each gondola is unique, some instruments are common to all three. Each gondola houses MEMS devices for analysis of gases and aerosols, a camera for visible and narrow UV band imaging, temperature and pressure sensors, an anemometer, and at least one tunable laser spectrometer.

Accommodation of the science instruments and other subsystems on all three gondolas requires some combination of protrusions, windows, atmospheric inlets, and outlets—all of which constitute sealed penetrations and present their own unique design challenges. Inlets and outlets must permit or force gases and aerosols through the pressure vessel wall in a controlled manner. Windows require the formation of leak-proof seals between two materials with vastly different properties (e.g., sapphire and titanium [21]) and must be accommodated to ensure mitigation of fogging. Protrusions, while less problematic than other types of sealed penetrations, still present disruptions in the pressure vessel structure and must be managed accordingly.

The purpose of Table 1 is to better elucidate the designs of the three gondolas, as well as the differences between them, namely those pertaining to instrument accommodation. A more comprehensive discussion of the instruments themselves and their scientific significance can be found in Section 4 and Appendix C of [34].

**Table 1.** VLF instrument summary. For a detailed discussion of the instruments and science behind the proposed VLF mission architecture, see [34].

| Instrument | Acronym | Number of Units | | |
| --- | --- | --- | --- | --- |
| | | Habitability | VAIHL Lite | VAIHL |
| Mini Tunable Laser Spectrometer | TLS | 1 | 3 | 1 |
| MEMS Gas Analyzer | MEMS-G | | 2 | 1 |
| MEMS Aerosol Analyzer | MEMS-A | | 2 | 1 |
| Nephelometer | NEP | | 1 | 1 |
| Autofluorescing Nephelometer | AFN | 1 | | |
| Mass Spectrometer | MS | | | 1 |
| Anemometer | AN | 1 | 1 | 1 |
| Solid Collector | SC | | | 1 |
| Liquid Collector | LC | | | 1 |
| Fluid-Screen | FS | | | 1 |
| Microscope | MP | | | 1 |
| Temperature and Pressure Sensor | TP | 1 | 1 | 1 |
| Camera | CAM | 1 | 1 | 1 |
| Antenna | ANT | 1 | 1 | 1 |

The inner radius of the VAIHL Mission gondola is 32.5 cm. Its mass is 145.3 kg. The hull and instrument shelf are each 6 mm thick. There are 16 sealed penetrations in the pressure vessel: five inlets provide direct atmospheric access to the TLS, MEMS-G, MEMS-A, LC, and SC; six outlets release sampled material back into the atmosphere; three protrusions accommodate the TP and AN; two windows provide optical access to the NEP and CAM (See Figure 8 and Table 2).

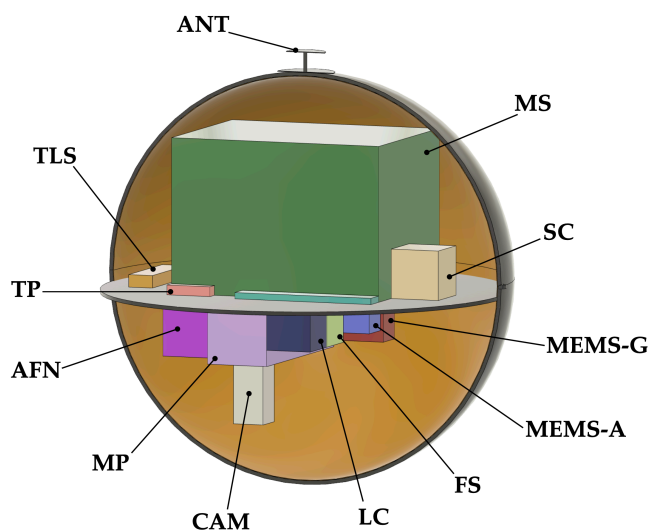

**Figure 7.** A schematic of a gondola for the VAIHL Mission. Non-science subsystems, except for the antenna, are not shown.

**Table 2.** VAIHL Mission Gondola mass and power breakdown.

| Component | Mass (kg) | | | Power (W) | | |
|---|---|---|---|---|---|---|
| | CBE | Cont. % | MEV | CBE | Cont. % | MEV |
| Structure | 37.4 | 30 | 48.6 | N/A | N/A | N/A |
| Science Instruments | 30.0 | 30 | 39.0 | 10.0 | 30 | 13.0 |
| Battery + PDS | 30.0 | 30 | 39.0 | 1.0 | 30 | 1.3 |
| Communication | 3.7 | 30 | 4.8 | 2.5 | 30 | 3.3 |
| Thermal | 7.5 | 30 | 9.8 | N/A | N/A | N/A |
| C&DH | 3.1 | 30 | 4.0 | 5.0 | 30 | 6.5 |
| **Total Payload** | **111.8** | **30** | **145.2** | **18.5** | **30** | **24.1** |

**CBE** = current best estimate; **Cont.** = contingency; **MEV** = maximum expected value; **PDS** = power distribution system; **C&DH** = command and data handling.

The inner radius of the VAIHL Lite Mission gondola is 15 cm. Its mass is 43.4 kg. The hull and instrument shelf are 3 mm thick. There are 11 sealed penetrations in the pressure vessel: three inlets provide direct atmospheric access to the TLS, MEMS-G, and MEMS-A devices; three outlets release sampled material back into the atmosphere; three protrusions accommodate the TP, AN, and ANT; two windows provide optical access to the NEP and CAM (See Figure 9 and Table 3).

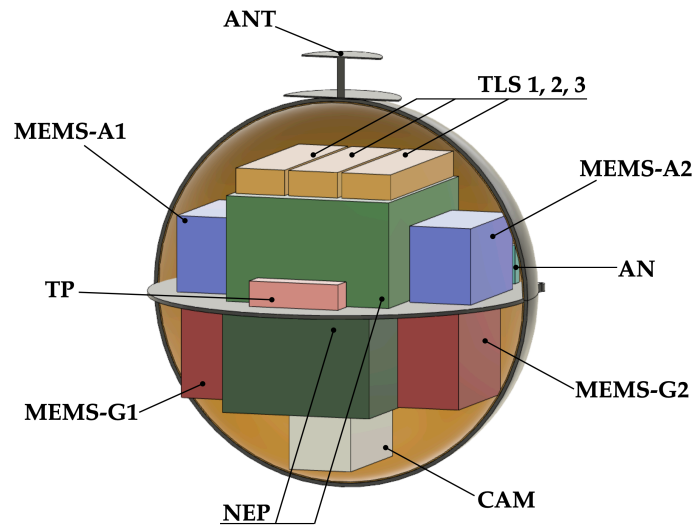

**Figure 8.** A schematic of a gondola for the VAIHL Lite Mission. Non-science subsystems, except for the antenna, are not shown.

**Table 3.** VAIHL Lite Mission gondola mass and power breakdown.

| Component | Mass (kg) | | | Power (W) | | |
| --- | --- | --- | --- | --- | --- | --- |
| | CBE | Cont. % | MEV | CBE | Cont. % | MEV |
| Structure | 4.3 | 30 | 5.6 | N/A | N/A | N/A |
| Science Instruments | 10.0 | 30 | 13.0 | 10.8 | 30 | 14.0 |
| Battery + PDS | 10.0 | 30 | 13.0 | 1.0 | 30 | 1.3 |
| Communication | 3.7 | 30 | 4.8 | 2.5 | 30 | 3.3 |
| Thermal | 2.3 | 30 | 3 | N/A | N/A | N/A |
| C&DH | 3.1 | 30 | 4.0 | 5.0 | 30 | 6.5 |
| **Total Payload** | **33.4** | **30** | **43.4** | **19.3** | **30** | **25.1** |

**CBE** = current best estimate; **Cont.** = contingency; **MEV** = maximum expected value; **PDS** = power distribution system; **C&DH** = command and data handling.

The inner radius of the Habitability Mission gondola is 10 cm. Its mass is 23.3 kg. The hull and instrument shelf are 2 mm thick. There are eight sealed penetrations in the pressure vessel: two inlets provide direct atmospheric access to the AFN and TLS; two outlets release sampled material back into the atmosphere; three protrusions accommodate the TP, AN, and ANT; a single window provides optical access to the CAM (See Figure 10 and Table 4).

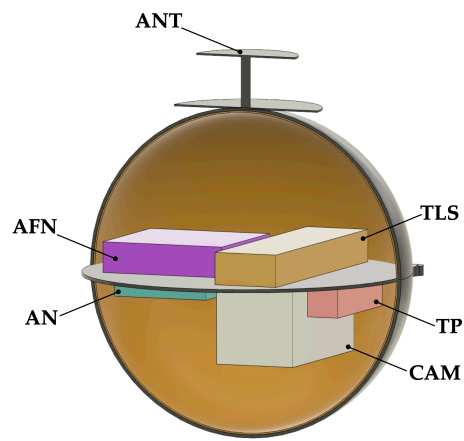

**Figure 9.** A schematic of a gondola for the Habitability Mission. Non-science subsystems, except for the antenna, are not shown.

**Table 4.** Habitability Mission Gondola mass and power breakdown.

| Component | Mass (kg) | | | Power (W) | | |
|---|---|---|---|---|---|---|
| | CBE | Cont. % | MEV | CBE | Cont. % | MEV |
| Structure | 1.3 | 30 | 1.7 | N/A | N/A | N/A |
| Science Instruments | 4.0 | 30 | 5.2 | 13.7 | 30 | 17.8 |
| Battery + PDS | 5.0 | 30 | 6.5 | 1.0 | 30 | 1.3 |
| Communication | 3.7 | 30 | 4.8 | 2.5 | 30 | 3.3 |
| Thermal | 0.9 | 30 | 1.2 | N/A | N/A | N/A |
| C&DH | 3.1 | 30 | 4.0 | 5.0 | 30 | 6.5 |
| **Total Payload** | **17.9** | **30** | **23.4** | **22.2** | **30** | **28.9** |

**CBE** = current best estimate; **Cont.** = contingency; **MEV** = maximum expected value; **PDS** = power distribution system; **C&DH** = command and data handling.

The differences between the three gondolas are attributable to their unique instrument suites. While the VAIHL Mission offers the most science return of the three options, the mass and volume of its gondola are considerably higher than the other candidates' and therefore require a larger balloon system and entry capsule.

The VAIHL Lite Mission features a subset of the instruments present in the suite of its larger counterpart. Even the tunable laser spectrometer is a reduced version of that featured in the VAIHL Mission. The Habitability Mission gondola omits further instrumentation, reducing the mass of its gondola to approximately 46% that of the VAIHL Lite Mission. To increase the science return of the Habitability Mission, it is proposed that a small number of mini probes be deployed at various times after the balloon reaches its initial float altitude (in a manner similar to that described in [36]). The mini probes search for the presence of metals and measure vertical profiles of selected gases and single droplet acidity. Data are transmitted to the gondola as the mini probes descend through the cloud decks [34].

### 7. Concepts of Operations

The entry, descent, and inflation (EDI) concepts of operations are essentially equivalent for all types of balloons—CAB or VAB. Figure 11 shows a typical sequence of events for EDI for a CAB. The probe entry interface is assumed at 180 km. After entry at approximately 11.3 km/s, the probe reaches peak deceleration and a stagnation heat rate between 80 and 90 km. The drogue chute is deployed at around 74 km, after which the heat shield is jettisoned and the aerial platform and inflation system are deployed.

Upon jettisoning the heat shield, the descent chute is deployed, and inflation begins. The rate of inflation is different for different balloon types, but full inflation occurs at no lower than 52 km. The balloon can operate nominally above 45 km, so this provides a margin of safety in deployment operations. After the envelope is fully inflated, the inflation system is jettisoned at a desired altitude between 52 and 50 km, and the balloon rises to its float altitude to begin science operations.

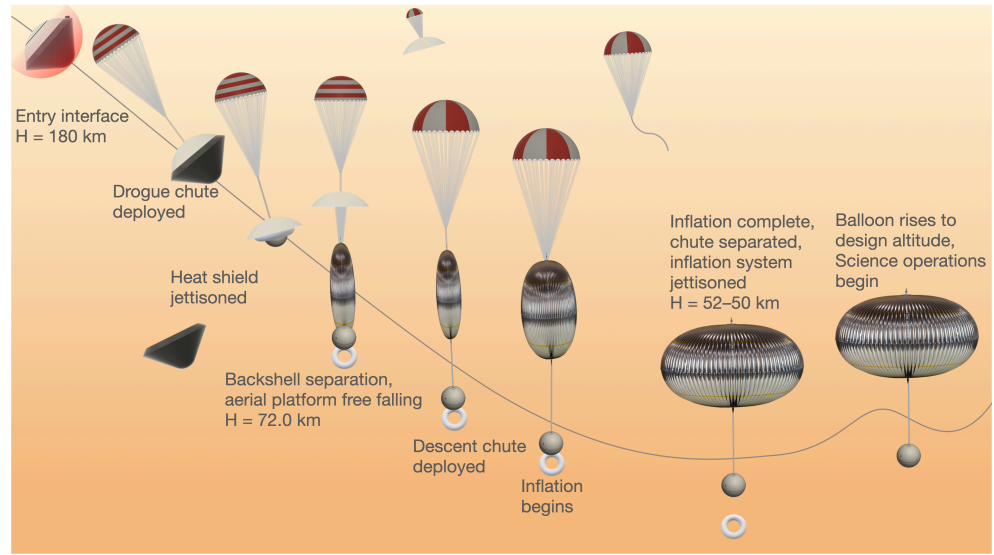

**Figure 10.** Representative concept of operations for a CAB.

## 8. Conclusions

Science has motivated the development of new atmospheric mission concepts at Venus. Extended in situ analysis of cloud material at multiple altitudes is key to the success of such missions. Existing analyses suggest that balloons, particularly those with altitude control capabilities, currently offer the highest ratio of science return to imposed risk. We consider a design option for a mechanical compression balloon capable of accessing various altitudes in the lower and middle clouds.

Constant-altitude balloons, while offering less science return, still present a valuable alternative and even have a flight heritage on Venus. We consider further design options—varying in scale—for CABs to access altitudes of interest in the lower and middle clouds.

Gondola design options appropriate for either balloon system are presented for atmospheric sampling missions of varying scope. Gondolas with more elaborate instrument suites are inevitably more costly in terms of mass and volume. However, for a given payload and all design options herein considered, the altitude control capability of the balloon imposes the harshest penalty on the size of the system.

The baseline architecture of the VLF mission proposes a notional launch date of 30 July 2026, which would place the cruise stage on Venus on approximately November 29 of the same year. Currently, the maturity of required science instruments and uncertainties in data volume and data rate capabilities limit the resolution of some areas of the mission design. Further development in these areas will help drive the final selection of an aerial platform suitable for a life-finding mission on Venus.

**Author Contributions:** Conceptualization, S.S. and S.J.S.; methodology, M.d.J.; software, M.d.J. and A.A.; formal analysis, W.P.B., M.d.J., R.A., and A.A.; investigation, W.P.B., M.d.J., R.A., and J.J.P.; resources, S.J.S.; data curation; writing—original draft preparation, W.P.B., M.d.J., R.A., and J.J.P.; writing—review, and editing M.d.J., R.A., J.J.P., S.J.S., S.S., and J.L.; visualization, W.P.B., M.d.J., and R.A.; supervision, S.S. and S.J.S.; project administration, S.S. and S.J.S.; funding acquisition, S.S. All authors have read and agreed to the published version of the manuscript.

**Funding:** This research was funded by Breakthrough Initiatives.

**Institutional Review Board Statement:** Not applicable.

**Informed Consent Statement:** Not applicable.

**Data Availability Statement:** Not applicable.

**Acknowledgments:** The authors would like to thank the members of the Venus Life Finder Mission team, the names of whom can be found at the following link: (hats://venuscloudlife.com/, accessed on July 3, 2022). The authors would also like to thank Breakthrough Initiatives for their partial funding of this study.

**Conflicts of Interest:** The authors declare no conflict of interest.